# Temperature Gated Thermal Rectifier for Active Heat Flow Control

Jia Zhu[1]*, Kedar Hippalgaonkar[1]*, Sheng Shen[1], Kevin Wang[2], Junqiao Wu[2], Xiaobo Yin[1], Arun Majumdar[1]*, Xiang Zhang[1]*

[1]Department of Mechanical Engineering, University of California at Berkeley, Berkeley, CA 94720, USA.
[2]Department of Materials Science and Engineering, University of California at Berkeley, Berkeley, CA 94720, USA.

*To whom correspondence should be addressed. E-mail: xiang@berkeley.edu and 4majumdar@gmail.com

Active heat flow control is essential for broad applications of heating, cooling and energy conversion. Like electronic devices developed for the control of electric power, it is very desirable to develop advanced all-thermal solid-state devices that actively control heat flow, without consuming other forms of energy. Here we demonstrate temperature-gated thermal rectification using vanadium dioxide beams, in which the environment temperature actively modulates asymmetric heat flow. In this three terminal device, there are two switchable states, which can be regulated by global heating. In the "Rectifier" state, we observed up to 28% thermal rectification. In the "Resistor" state, the thermal rectification is significantly suppressed (<1%). To the best of our knowledge, this is the first demonstration of solid-state active-thermal devices, with a large rectification in the "Rectifier" state. This temperature-gated rectifier can have substantial implications ranging from autonomous thermal management of heating and cooling systems to efficient thermal energy conversion and storage.

---

* These authors contributed equally to this paper.

Heat and charge transport in condensed matter were first characterized about two centuries ago by the well-known Fourier's [1] ($\vec{q} = -k\nabla\vec{T}$) and Ohm's [2] ($\vec{J} = s\vec{E}$) laws, respectively. The history of how the science of heat and charge transport has evolved, however, is very different. The progress in material processing (such as purification of semiconductors) and fundamental understanding (quantum mechanics) led to the invention of many electronic devices such as transistors to actively control and manipulate charge transport. Such devices have been widely deployed, and have touched almost all aspects of modern life in what we now call the information revolution. Heat transport in condensed matter, in stark contrast, has remained in the realm of the Fourier law and its manipulation beyond has been largely absent. Yet, about 90 percent of the world's energy utilization occurs through heating and cooling, making it one of the most critical aspects of any modern economy [3]. Hence, the ability to actively manipulate heat transport in ways akin to that for charge transport could potentially significantly impact utilization of energy resources.

A few theoretical proposals have been made envisioning control of heat flow in solid-state devices [4,5] and electrically tuned solid-state thermal memory has recently been experimentally realized [6]. Most development has explored the possibility of thermal rectification [7,8], in which the system thermal conductance depends on the direction of thermal gradient. The level of thermal rectification is commonly [7] defined as the following:

$$R = \frac{G_H - G_L}{G_L}$$

Here $G_H$ and $G_L$ are the thermal conductances of the sample in the directions of higher and lower heat flows under the same temperature difference, respectively. Several approaches have been theorized for achieving thermal rectification, such as using materials with opposite trends in thermal conductivity as a function of temperature[9,10], or asymmetrical phonon density of states in graphene nanoribbons [11]. The rapid advancement of nanofabrication has enabled the synthesis of nanostructures with a variety of materials for both novel applications[12–16] and to explore condensed matter science[17]. Specifically, individual carbon or boron nitride nanotubes with asymmetric mass loading were reported to have thermal rectification of about 2-7% [7]. However, a pure active thermal device where heat as an input can modulate thermal transport has never been realized.

In condensed matter, both phonons and electrons carry heat. Metals are generally good conductors of heat through electrons. Manipulating heat in non-metals requires tuning of quantized lattice vibrations or phonons. Heat transfer with a single energy carrier (either electrons and phonons) has been studied in detail in many condensed matter systems. Can the interplay between the two energy carriers at metal-insulator interfaces potentially lead to an asymmetry when the direction of heat flow is reversed [8]? Very early work in a CuO-Cu system showed thermal rectification due to a single metal-insulator interface [18]. Here we demonstrate the first temperature-gated thermal rectifier devices using $VO_2$ beams. Interestingly, thermal rectification in the beams can be actively switched on and off by changing the device temperature, which controls the metallic and insulating phases, thus functioning as a gate. Rectification up to 28% is observed below 340 K. Once the devices are heated above 340 K, they can be switched off, where thermal rectification is greatly suppressed (<1%) ( see Fig. 1A for a depiction of the device functionality).

Our material system of choice for this study is the family of vanadium oxides. Single crystalline $VO_2$ beams have been investigated extensively as a unique material system for studying the complexity of metallic and insulating phases which can be induced by temperature, strain, stoichiometry and light [19–23]. Thin films of polycrystalline $VO_2$ have shown that the insulator-to-metal transition occurs at the transition temperature ~340K via nucleation of isolated nanoscale puddles of metallic phases in a background of the insulator phase, which then grow and merge as the transition progresses [24]. The thermal conductivity of poly-crystalline stoichiometric $VO_2$ films was also studied close to ~340K and increased by as much as 60% due to the phase transition [26].

We synthesized $VO_2$ beams using a modified vapor transport method (Methods), and studied their electrical properties as well as overall structure with Raman spectroscopy (more details in Supplementary Information 1) [27]. The structure and electrical properties confirm a $VO_2$ backbone. An individual 5-10 μm long $VO_2$ beam is then transferred to a silicon microdevice so as to form a bridge between two parallel, suspended $SiN_X$ membranes, each consisting of micro-fabricated symmetric resistive platinum coils, for thermal and electrical transport measurements [28] (Fig. 1B). The platinum coils are used as both heaters and resistive thermometers (Methods). To make good electrical and thermal contact, a platinum/carbon composite was deposited symmetrically on both ends using a focused ion beam. Therefore, both thermal conductance and electrical conductance can be measured at the same temperature. A resistive heater is used to heat the whole chip uniformly inside a cryostat to control the global device temperature, $T_G$. The measurement is performed in a cryostat at ~2 µTorr to prevent conduction and convection losses. (Supplementary Information 2).

Figure 2A shows measured heat flow, $Q$, and the temperature difference, $\Delta T$, across a tapered beam at $T_G$=320K below the VO$_2$ insulator-metal phase transition temperature of 340 K. The heat flow, $Q$, increases linearly with $\Delta T$ (<1K). However, the thermal conductance, $G = Q/\Delta T$, or the slope, differs depending on the direction of heat flow. The thermal conductance of the tapered beam when heat is flowing from the narrow to the broader side (62.2 ± 0.5 nW/K) is significantly smaller than that in the other direction (80.1 ± 0.6 nW/K). This represents a 28 ± 1.4% thermal rectification, which is the highest ever reported to the best of our knowledge. Also shown in Fig. 2B is $\delta Q$, which is the deviation of the heat flow in one direction, $Q$, from the extrapolated linear curve representing the conductance in the opposite direction. Figures 2C and 2D show similar plots for the same VO$_2$ beam at $T_G$ = 385 K, which is higher than the phase transition temperature where the VO$_2$ beam is in the metallic phase. In contrast to Figs. 2A and 2B, it is clear that no rectification is observed and the thermal conductance in both directions increases to 94 ± 0.4 nW/K. The same plots of $Q$ and $\delta Q$ at 320 K of a uniform or untapered VO$_2$ beam show that there is no rectification (Supplementary Information 3a).

The thermal rectifications of six asymmetrical VO$_2$ beams were found to be around 10-20%, when $T_G$ < 340K, and reduced to below 1% for $T_G$ > 340K (Table 1). Note, that in beam V, the rectification persists above the purported Insulator-to-Metal transition for stoichiometric VO$_2$. We ascribe this to an excess of oxygen, thus shifting the transition temperature to much higher temperatures. As control experiments, symmetric beams with uniform width, which have no variation in stoichiometry, are also measured and no apparent thermal rectification effects are observed (<3%) at all temperatures below and above the transition temperature of 340K (Supplementary Information 3b). To accurately measure the thermal conductance of $VO_2$ beams, three different methods are used all with consistent results (Supplementary Information 4).

Figure 3A shows the thermal conductance in the two directions (green and red) as a function of global temperature for a representative $VO_2$ beam. Between 250K and 340K, the conductance is different in different directions, which is discussed in detail later. It is observed that the degree of rectification, *R*, (black open circles) peaks at 320K, lower than the phase transition of 340K and decreases as the temperature is increased or decreased away from the transition. Above the $VO_2$ insulator-metal transition temperature (~340K), the electronic density of states at the Fermi Level increase significantly and electrons start contributing to the thermal conductance, which explains a sudden increase. This is also consistent with electrical resistance measurements (see Figure 3B), which show a drop in magnitude by two orders at 340K indicating the characteristic Insulator-to-Metal phase transition, consistent with reports in literature for the electronic $VO_2$ phase transition [22]. Below 135K, the electronic contribution is negligible and phonons dominate heat conduction. In this temperature range, we do not observe any rectification as well. The phonon mean free path is limited by scattering from either defects, interfaces or boundaries, and the thermal conductance increases with temperature due to increase in phonon population. Temperature dependent conductance and rectification plots similar to Figure 3A for beam II are shown in Supplementary Information 5.

In order to understand the mechanism behind temperature dependent rectification, we first consider the effect of the asymmetric geometry on the phonon mean free path. It has been proposed that an asymmetric geometry or roughness may cause thermal rectification in materials when phonons dominate heat conduction and the phonon mean free path is comparable to the characteristic length of the structure [4]. However, the phonon mean free path in single crystal $VO_2$ is estimated to be of the order of a few nanometers at room temperature (about 300 K) in both the metallic and insulating phases [29,30] (Supplementary Information 6), significantly smaller than the scale of the beams employed in this study (300 nm to a few microns). Below 50 K, the phonon mean free path should increase by at least one order of magnitude. Therefore, any thermal rectification caused by asymmetrical geometry should be a lot more significant at lower temperature, which is not observed in Figure 3A. So it is not possible that uneven phonon heat conduction due to asymmetrical geometry could cause the observed large thermal rectification. This further indicates that the macroscopic geometric ratio between the broad and narrow widths of the beam across the taper is not expected to scale with the observed thermal rectification.

Next, we explore how the Insulator-to-Metal phase transition in $VO_2$ can be responsible for large thermal rectification observed here. Previous work has shown the co-existence of metallic and insulating phases within a single beam during phase transition [19,20,22]. A recent theoretical study estimated the thermal resistance between a metal and an insulator by employing the two-temperature model to account for the lack of equilibrium between electrons and phonons near a metal-insulator interface [31] while maintaining the same lattice temperature. It has been further theoretically predicted that thermal rectification could occur if metallic and insulating phases co-exist in a material system [8]. The underlying principle is the asymmetry between the energy transfer rate between electrons and phonons, when the electrons and phonons are not in equilibrium in a metal close to the interface. To observe significant thermal rectification in a metal-insulator system, the thermal resistance due to electron-phonon scattering should dominate over the phonon-phonon coupling resistance. Indeed, in the vanadium oxide system, a small ~1% lattice distortion [22] in the rutile and monoclinic phases should ensure good acoustic match to reduce thermal resistance from phonon-phonon coupling. Therefore, it is possible that the electron-phonon scattering may be dominant for the thermal resistance at metal-insulator interfaces in $VO_2$ beams.

For measurable thermal rectification, an abundance of co-existing metallic and insulating phases must prevail in the beams. Intriguingly, thermal rectification was observed not only near phase transition temperature 340K, but also over large span of ~100K below, provoking the question as to the role of the taper of the $VO_2$ beams. Even though the $VO_2$ beams show characteristic electronic transition temperature (340K) and Raman spectrum (Supplementary Information 7 and 1 respectively), it has been known that vanadium oxide can form Magnéli phases with a deficiency of oxygen, given by the general formula $V_nO_{2n-1}$, or excess of oxygen described by $V_nO_{2n+1}$. These are crystallographic shear compounds with a rutile $VO_2$ backbone [32]. The role of stoichiometry in $V_nO_{2n-1} = V_2O_3 + (n-2)VO_2$, or $V_nO_{2n+1} = V_2O_5 + (n-2)VO_2$ single crystals has been studied in meticulous detail [25]. As observed in Figure 3A, the 'on' state of rectification exists between 250K and 340K, where the $V_2O_3/V_2O_5$ shear planes would be metallic and the $VO_2$ matrix would be insulating. Therefore, a small variation in stoichiometry of vanadium oxide can cause the existence of metal-insulator interfaces over a very large range of temperature [25]. This is suggested in Figure 3A, between 250 K and 340K where rectification exists, in the regime where thermal conductance decreases with increasing temperature. While phonon-phonon Umklapp scattering is one possible cause for this behaviour, it can also be attributed to the appearance of interfaces created by the formation and co-existence of multiple phases of vanadium oxide that may be not electronically connected, nevertheless impeding phonons due to interface scattering.

We employed scattering type scanning near-field optical microscope (s-SNOM) to probe the possible existence of mixed phases. s-SNOM allows direct imaging of evolution of insulating and metallic phases with increasing temperature at nanometer-scale spatial resolution as demonstrated previously for microcrystals and polycrystalline films. A linearly polarized probing $CO_2$ laser (wavelength, $\lambda=10.7$ μm) is focused on the tip-sample interface. High harmonic demodulation coupled with pseudo-heterodyne interferometer are used to detect the near-field signal with ~30nm spatial resolution above the tapered $VO_2$ beam [24,33]. The image contrast is determined by the local spatially varying dielectric function of the surface. Hence, regions of the metallic phase, due to larger effective tip-sample polarizablility, result in higher s-SNOM amplitude signal compared with that of the insulating phase (Supplementary Information 8). The wavelength of the laser was chosen specifically in order to maximize the difference in the optical conductivity of the insulating and metallic phases of $VO_2$ and thus achieve optimum s-SNOM amplitude contrast at different temperatures. Interestingly, for a tapered beam (AFM topography shown in Fig 4A) demonstrating ~15% thermal rectification at 300K (data not shown in Table 1 as device broke before reaching 340K during measurement), there exists s-SNOM amplitude contrast across the taper confirming phase coexistence at the same temperature (Fig 4B), which disappears when the sample is heated up to 350K (Fig 4C). Note, that the s-SNOM images show on average that the narrow side of the tapered VO2 beam is metallic, while the broader side is insulating. Since the resolution is limited to ~30nm, we did not observe any crystallographic shear planes with this technique. A similar experiment on a uniform cross-section beam shows no amplitude contrast evolution (Supplementary Information 8).

While the s-SNOM figures (4A-C) show unequivocally the presence of mixed phases, the origin of these remains to be ascertained. As explained above, these could be due to a variation of stoichiometry along the beam length. We also found a signature of mixed vanadium oxide phases in tapered beams using Auger Electron Spectroscopy (AES) (Supplementary Information 9). In addition to stoichiometry, recent work shows that the phase transition also depends on stress fields within single $VO_2$ beams [22]. It is possible that the taper created during the beam growth may lead to stress gradients which could produce geometrical and size distributions of metallic and insulating domains and interfaces near the taper. These can amplify the rectification achieved by single interfaces. Similar distributions were previous reported by bending the beam [19]. Hence, while the Insulator-to-Metal phase transition is critical to thermal rectification, the taper and composition variation may also contribute to the effect by unique distributions of metal-insulator domains. Indeed, we also found that a uniform, but non-stoichiometric $VO_x$ beam with huge variation in stoichiometry across its length showed thermal rectification. Also, due to the excess of oxygen in this beam, it did not undergo the Insulator-to-Metal transition at 340K (Supplementary Information 10). Hence, we could not switch off the thermal rectification in this beam. The metal-insulator domains and interfaces can be rationally engineered to control the thermal transport. In stoichiometric VO$_2$ beams, an array of metal domains can be created below 340K by either substitutional doping [34,35] or local stressing [19,22]. In addition, metal domains can be stabilized along these VO$_2$ beams at sub-340K temperatures by encoding stoichiometry variation during the growth [21] or post-growth hydrogenation [36]. Over the last several decades, while tremendous progress has been made in understanding the complexity of the phase transition in the family of vanadium oxides, the underlying physics still remains largely illusive [37]. Its impact on thermal transport is much less explored and has yet to be fully understood. Rational synthesis of vanadium oxide based beams with controlled local stresses and/or stoichiometry with a lack or excess of oxygen can open up pathways to further manipulate heat transfer in these systems.

In summary, we report a large thermal rectification up to 28% in VO$_2$ beams that is gated by the environmental temperature. It is the first demonstration of an active three-terminal thermal device exhibiting an "on" rectifying state over large range of temperature ($T_G$ = 250-340 K) and "off" resistor state ($T_G$ < 250 K or $T_G$ > 340K). By changing temperature, one can switch the rectification, much like a gate voltage switches a thyristor between two states of electrical conductance. The realization of such unique thermal control is a consequence of the interplay between metallic and insulating phases due to the rich parameter space provided by the vanadium-oxygen material family. Such novel all-thermal devices may spurn interesting applications in autonomous thermal flow control and efficient energy harvesting.

**Methods**

Synthesis of VO$_2$ beams

Bulk VO$_2$ powder was placed in a quartz boat in the center of a horizontal tube furnace. The typical growth temperature was 1000 ºC with Ar used as the carrier gas. The VO$_2$ beams were collected on a Si substrate with a 500nm thick thermally grown surface oxide downstream from the source boat. The catalyst, which determines the size of the beam, can be partially diffused away by tuning the pressure and temperature to induce tapered or asymmetrical beam growth.

**Thermal Conductance Measurement of Single VO$_2$ Beam**

A resistive heater is used to heat the whole Si chip uniformly inside a cryostat to control the global device temperature, $T_G$. For thermal conductance measurements, a small DC current (~6 µA) is passed through the Pt coil on one of the membranes to heat it to a temperature, $T_h$, above $T_G$, thus inducing a heat flow Q through the VO$_2$ beam to heat up the other membrane to $T_s$. An AC current of 500 nA is passed through the Pt coils on both membranes to determine its electrical resistance through a 4-point technique, which is then used to estimate the temperatures $T_h$ and $T_s$. Using two SRS 850 lock-in amplifiers for the AC signals, signals from the sensing side were measured using a frequency of 199 Hz whereas that for the heating side utilized 1.11 kHz (more details in Supplementary Information 3). Also, the effect of contact resistance on thermal rectification is discussed in Supplementary Information 11. Finally, error analysis is in Supplementary Information 12.

**Acknowledgements**

This materials synthesis part was supported by the U.S. Department of Energy Early Career Award DE-FG02-11ER46796.

Figure 1

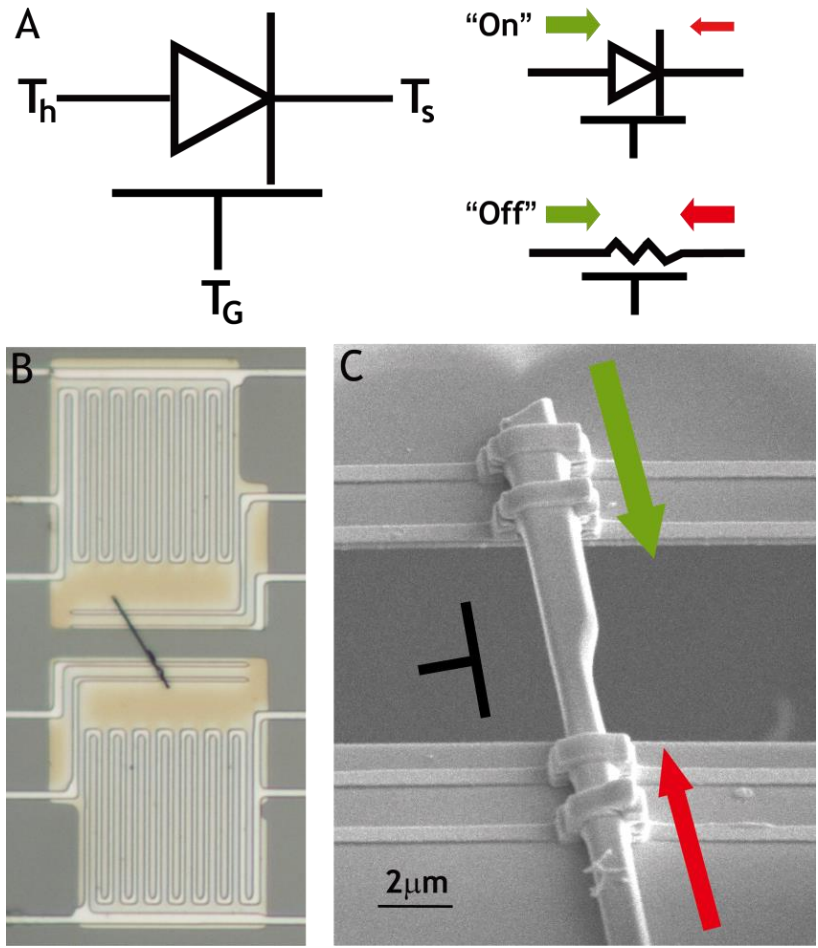

Fig. 1 A) Symbolic diagram of temperature-gated thermal rectifier. In the "Rectification" or "on" state, thermal flow depends on the direction of applied thermal gradient, representing thermal rectification. In the "Resistor" or "off" state, thermal flow does not depend on the sign of thermal gradient, essentially the behavior of a resistor. The on/off state can be controlled by a global temperature ($T_G$). B) Optical microscope image of an asymmetrical $VO_2$ beam on suspended membranes for thermal conductance measurement. C) Scanning electron microscopy (SEM) image of an asymmetrical $VO_2$ beam. The $VO_2$ beams used in this study have a uniform thickness (typically 500nm-1μm), with one end of narrow width (300nm- 900nm) and the other end of wide width

(600nm - 2µm). The heat flow through the beam ($Q$) in either direction denoted by the arrows is accurately measured while the suspended platforms are maintained as isotherms at hot and cold temperatures, $T_h$ and $T_s$ respectively (details in Supplementary Information).

Figure 2

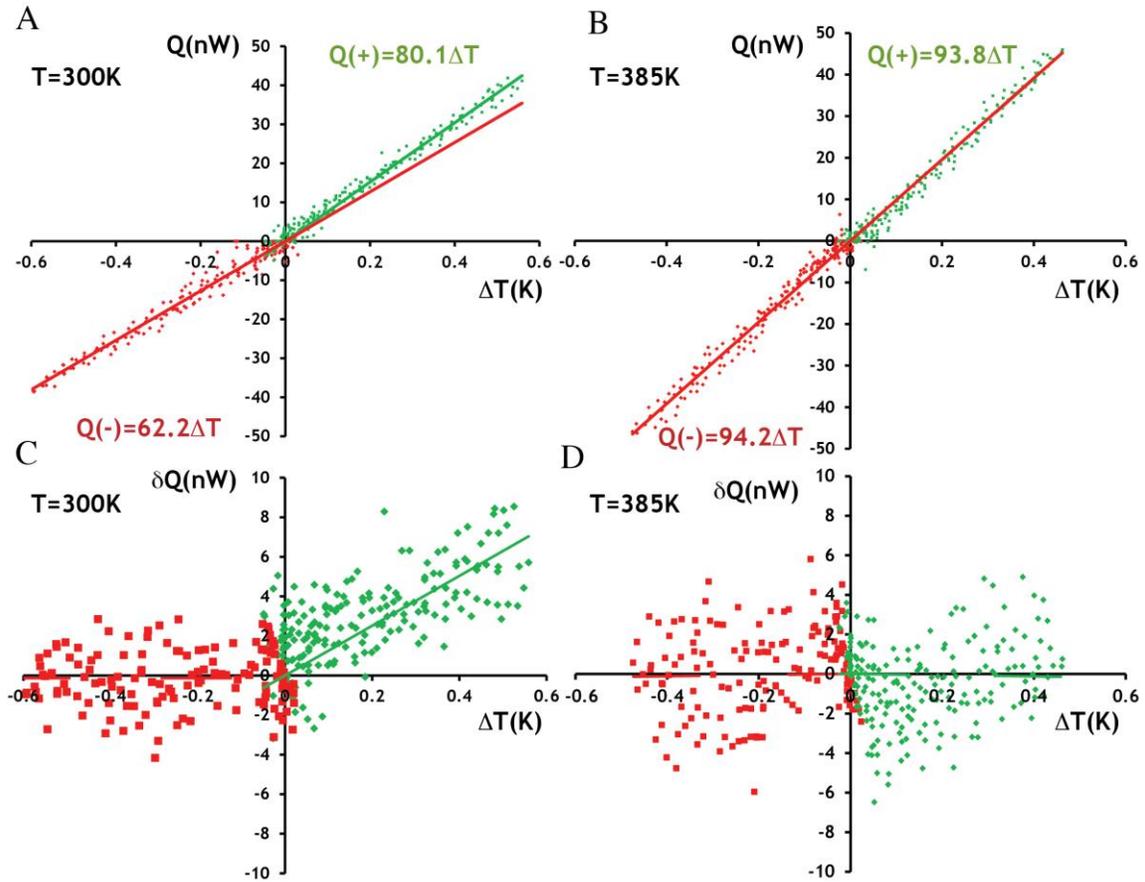

Fig. 2 A) and C) heat flow (Q) as a function of temperature difference ΔT across the VO$_2$ beams at 300K and 350K, respectively. Different signs (+) and (-) of Q represent different directions of heat transfer. B) and D) heat flow deviation (δQ) as a function of temperature difference across the VO$_2$ beams at 300K and 385K, respectively.

Figure 3

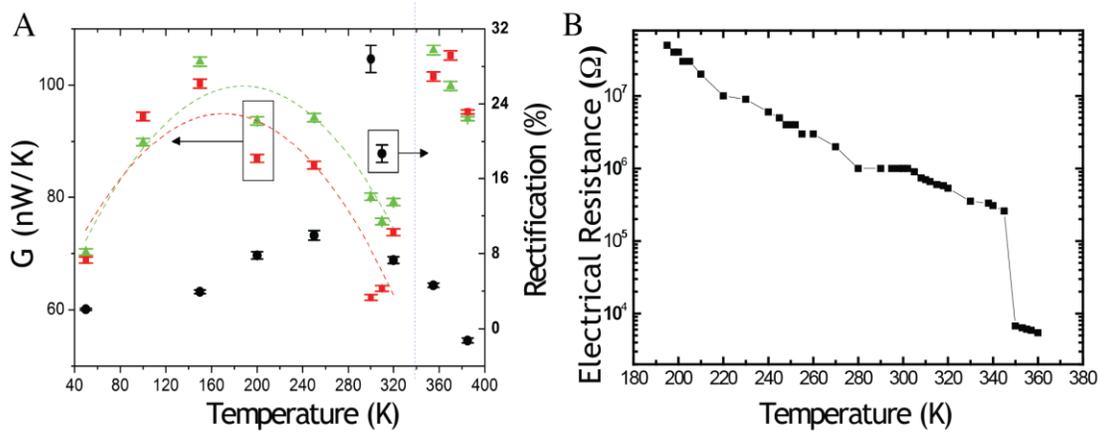

Fig. 3 A) Thermal conductance of an asymmetrical $VO_2$ beam as a function of global temperatures along two opposite directions (green triangles and red squares). The thermal conductance is found to be measurably higher when heat flows from the wide end to the narrow end. At low temperatures, the rectification (open black circles) disappears and the conductance from either end is identical; this is expected as the whole wire is in the insulating phase and should behave as a normal dielectric. B) the electrical resistance of $VO_2$ beam as a function of global temperature. The resistance could not be measured below 180K due to saturation. The arrows in Figure 1 denote the direction of heat flow in which high (green) and low (red) thermal conductance was observed.

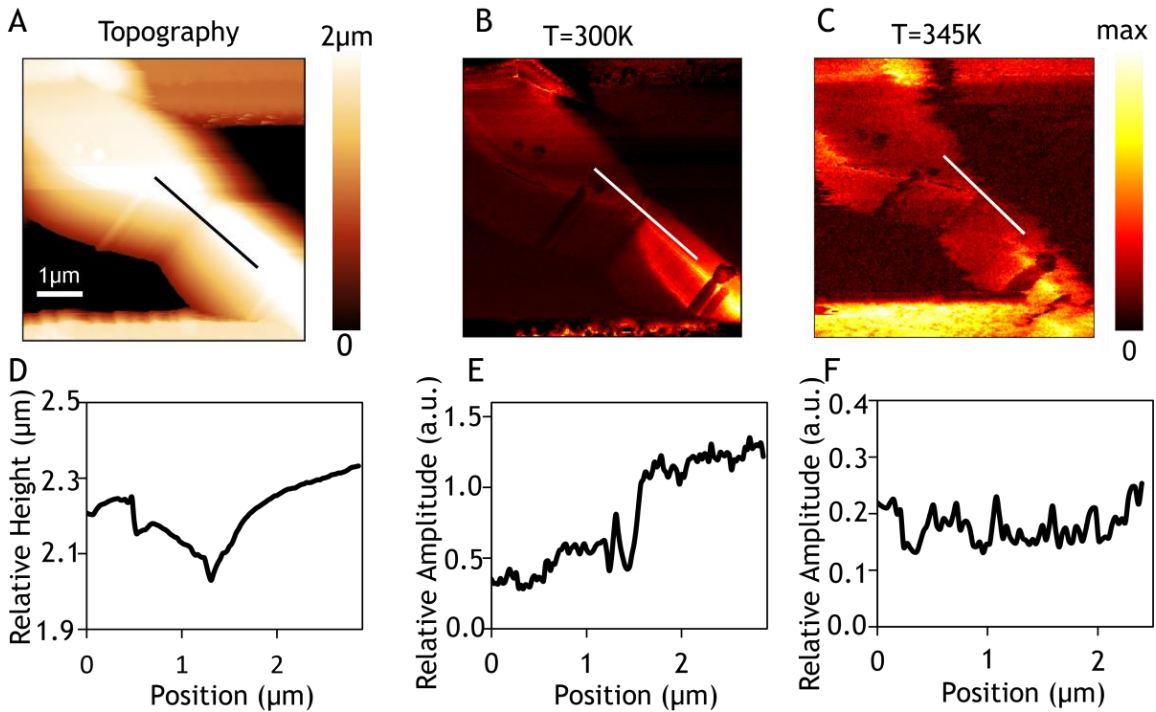

Fig. 4 (A) and (D) AFM topography obtained via tapping mode on a suspended tapered VO$_2$ beam exhibiting 15% rectification at 300K. The edges of the suspended membranes holding the beam are visible on the top and bottom of the image. A line profile is drawn across the tapered region on the beam, with the relative height in μm illustrated in (D) as a function of beam length. (B) Near-field amplitude signal obtained at second harmonic demodulation at 300K on the same tapered VO$_2$ beam. The black spots on the beam are dirt particles, possibly carbon, with a negligible dielectric value at the excitation frequency corresponding to a CO$_2$ laser wavelength of 10.7 μm. A white line profile is drawn across the taper, with the corresponding line profile of the near field signal amplitude in (E). The near-field amplitude contrast changes significantly across the taper, with the narrow region of the beam being more metallic and the broad section insulating. (C) The near field amplitude signal obtained on the same tapered VO$_2$ beam at 345K, higher than the insulator-metal phase transition temperature of VO$_2$. The black

spots, which are impurities on the surface, remain dark in contrast indicating temperature independent contrast. A white line is drawn at the same location as that in (B) across the taper, with its corresponding near field amplitude line profile illustrated in (F). The metal-insulator contrast difference seen in (E) at 300K across the taper disappears at this higher temperature of 345K, illustrating that the phase transition has occurred and that the phase is homogeneously metallic.

| No. | $T_G < 340K$ | | | $T_G > 340K$ | | |
|---|---|---|---|---|---|---|
| | G+ (nW/K) | G- (nW/K) | R (%) | G+ (nW/K) | G- (nW/K) | R (%) |
| I | 80.1±0.6 | 62.2±0.5 | 28.8±1.4% | 93.8±0.4 | 94.2±0.4 | -0.5±0.6% |
| II | 144±2.1 | 123±1.6 | 16.9±2.0% | 165±2.7 | 167±3.0 | 1.0±2.4% |
| III | 113±0.3 | 103±0.3 | 9.1±0.4% | 119±0.4 | 117±0.3 | 1.1±0.4% |
| IV | 191±0.8 | 171±0.6 | 12.1±0.6% | 181±0.6 | 181±0.6 | 0.0±0.5% |
| V | 58.5±0.1 | 55.3±0.1 | 5.9±0.2 | 58.6±0.1 | 56.8±0.1 | 3.2±0.2 |
| VI | 181±0.5 | 169±0.6 | 7.2±0.5% | 186±0.5 | 184±0.6 | 0.9±0.5% |

Table 1 Thermal conductance and maximum thermal rectification (at $T_G$=300-320K) of six different $VO_2$ beams